\documentclass[prl,nofootinbib,twocolumn,superscriptaddress]{revtex4}
\usepackage{amsfonts}
\usepackage{amsmath}
\usepackage{amssymb}
\usepackage{amsthm}
\usepackage{mathrsfs}
\usepackage{bm}
\usepackage{cancel}
\usepackage{graphicx}
\usepackage{hyperref}
\usepackage[usenames,dvipsnames]{color}

\newcommand{\st}[1]{\text{\tiny \rm #1}}

\def\e{\emph}

\def\shs{{\sf S}}
\def\pshs{{\sf PS}}

\def\bq{\begin{equation}}
\def\ee{\end{equation}}

\hypersetup{
   colorlinks=true,         
   linkcolor=blue,          
   citecolor=red,        
   urlcolor=Violet            
}

\begin{document}

\title{Identification of a gravitational arrow of time}

\author{Julian Barbour}
\affiliation{College Farm, South Newington, Banbury, Oxon, OX15 4JG UK,\\
Visiting Professor in Physics at the University of Oxford, UK.}

\author{Tim Koslowski}
\affiliation{University of New Brunswick, Fredericton, NB, E3B 5A3 Canada.}

\author{Flavio Mercati}
\affiliation{Perimeter Institute for Theoretical Physics, 31 Caroline Street North,\\
\it \small Waterloo, ON, N2L 2Y5 Canada.}

\begin{abstract}
It is widely believed that special initial conditions must be imposed on any time-symmetric law if its solutions are to exhibit behavior of any kind that defines an `arrow of time'. We show that this is not so. The simplest non-trivial time-symmetric law that can be used to model a dynamically closed universe is the Newtonian $N$-body problem with vanishing total energy and angular momentum. Because of special properties of this system (likely to be shared by any law of the Universe), its typical solutions all divide at a uniquely defined point into two halves. In each a well-defined measure of shape complexity fluctuates but grows irreversibly between rising bounds from that point. Structures that store dynamical information are created as the complexity grows and act as `records'. Each solution can be viewed as having a single past and two distinct futures emerging from it. Any internal observer must be in one half of the solution and will only be aware of the records of one branch and deduce a unique past and future direction from inspection of the available records.
\end{abstract}

\maketitle

Many different phenomena in the Universe are time-asymmetric and define an arrow of time that points in the same direction everywhere at all times \cite{zeh2007physical}. Attempts to explain how this arrow could arise from time-symmetric laws often invoke a `past hypothesis': the initial condition with which the Universe came into existence must have been very special. 
This is based on thermodynamic reasoning, which seems to make a spontaneous emergence of an arrow of time very unlikely. Although thermodynamics works very well for subsystems,
provided gravity is not a dominant force, self-gravitating systems exhibit
`anti-thermodynamic' behavior that is not fully understood. Since the
Universe is the ultimate self-gravitating system and since it cannot be
treated as any subsystem, its behavior may well confound
thermodynamic expectations.

In this Letter, we present a gravitational model in which this is the case. In all of its typical solutions, internal observers will find a manifest arrow of time, the nature of which we are able to precisely characterize.
We emphasize that in this letter we make no claim to explain all the various 
arrows of time. We are making just one point: an arrow of time does arise in at 
least one case without any special initial condition, which may therefore be 
dispensable for all the arrows. In this connection, we mention that in \cite{CarrollChen} (Section II) Carroll and Chen conjectured that the thermodynamic arrow of time might have a time-symmetric explanation through entropy arrows much like the complexity and information arrows we find.

\begin{figure*}[t]
\includegraphics[width=\textwidth]{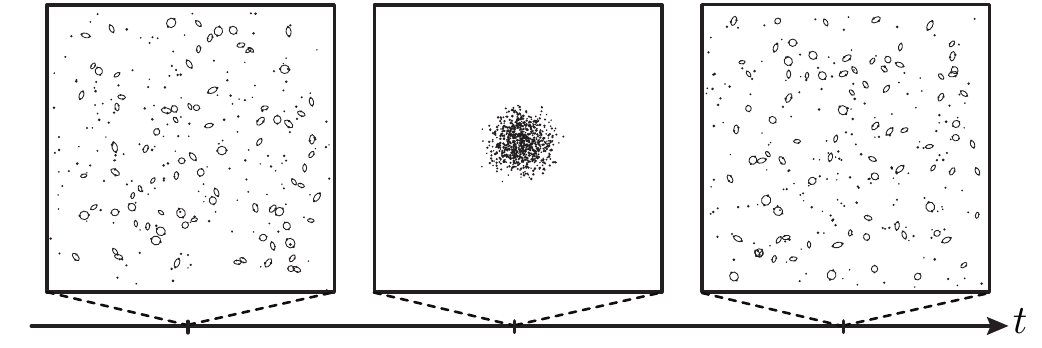}
\caption{Three configurations at different Newtonian times $t$ of a typical solution of the  $N$-body problem with ${\bf J}_\st{tot} =  0,\,E_\st{tot}=0$. 
The time symmetry of the law is reflected in qualitative symmetry about the central region in which the distribution of the particles is maximally uniform. The direction of time indicated by the arrow of the time axis is purely conventional. Either choice of direction gives contraction with structure destruction through uniformity at minimal size followed by expansion and structure formation, mainly in the form of Kepler pairs (shown as loops). Internal observers must be on one side of the central region and would regard it as their past.} \label{TriplePicture}
\end{figure*}

\section{The Model}

The Newtonian $N$-body problem  with vanishing total energy, $E_\st{tot}=0$, momentum, ${\bf P}_\st{tot}  = 0$, and angular momentum, ${\bf J}_\st{tot}=0$, is a useful model of the Universe in many respects \cite{GibbonsEllis}. As we show below, these conditions match the intution that only relational degrees of freedom of the Universe should have physical significance \cite{barbourbertotti:mach,JuliansReview,FlaviosSDtutorial}. A total angular momentum ${\bf J}_\st{tot}$ and a total energy $E_\st{tot}$ would define, respectively, an external frame in which the Universe is rotating and an absolute unit of time. Moreover the conditions ${\bf J}_\st{tot} = {\bf P}_\st{tot}  = 0,\,E_\st{tot}=0$ ensure scale-invariance and are close analogues of the Arnowitt--Deser--Misner constraints of Hamiltonian General Relativity (in the spatially closed case) \cite{LongPaper}. These properties, along with the attractivity of gravity, are architectonic and likely to be shared by any fundamental law of the Universe.

First support for our claim follows from the rigorous results of \cite{Marchial:1976fi} that asymptotically (as the Newtonian time $t\rightarrow \pm\infty$) $N$-body solutions with $E\st{tot}={\bf J}_\st{tot} =0$ typically `evaporate' into subsystems (sets of particles with separations bounded by $\mathcal O(t^{\frac 2 3})$) whose centers of mass separate linearly with $t$ as $t\rightarrow\pm\infty$.\footnote{The separation linear in $t$ is, of course, a manifestation of Newton's 1st law in the asymptotic regime. Our result arises from the combination of this behavior with gravity's role in creating subsystems that mostly then become stably bound.} Each subsystem consists of individual particles and/or clusters whose constituents remain close to each other, i.e., the distances between constituent particles of a cluster are bounded by a constant for all times. One finds in numerical simulations that the bulk of the clusters are 2-body `Kepler pairs' whose motion asymptotes into elliptical Keplerian motion. For reasons we next discuss, this $t\rightarrow\pm\infty$ behavior occurs in all typical solutions either side of a uniquely defined point of minimum expansion, as illustrated in Figure~\ref{TriplePicture}.

We identify and explain the arrows of time in the $N$-body solutions through the way the {\it shape} degrees of freedom (defined below) evolve and now explain why.

\section{Elimination of Scale}

Let ${\bf r}_a$ denote the position of particle $a$, ${\bf p}^a$ its momentum and $m_a$ its mass. We consider the Newtonian $N$-body problem with vanishing total energy $E_\st{tot} = 0$:
\begin{equation}
E_\st{tot} = \sum_{a =1}^N \frac{\mathbf p^a \cdot \mathbf p^a  }{2 \, m_a} +   V_\st{New} ,\,\,\,   V_\st{New} = - \sum_{a<b}{m_a m_b\over r_{ab} } \,,
\end{equation}
where $r_{ab}:=||{\bf r}_a-{\bf r}_b||$. The dynamics generated by a potential $V_k$ homogeneous\footnote{A function $f(x_1,\dots,x_n)$ is homogeneous of degree $k$ if $f(\alpha x_1,\dots,\alpha x_n) = \alpha^k f(x_1,\dots,x_n)$ for all $\alpha >0$.} of degree $k$ satisfies \emph{dynamical similarity} \cite{Landau-Lifshitz}: the anisotropic rescaling 
\begin{equation}\label{DynamicalSimilarity}
{\bf r}_a \to \alpha ~ {\bf r}_a, ~~~ t \to \alpha^{1-k/2} t \,
\end{equation}
of spatial distances and Newtonian time sends solutions into solutions. Kepler's third law is a consequence of this property
in the particular case $k = -1$. Dynamical similarity enables us to eliminate scale from the $N$-body problem. As overall scale of the system
it is natural to use $I_\st{cm}$, the centre-of-mass moment of inertia,
\begin{equation}
  I_\st{cm} :=  \sum_{a=1}^N m_a \|  {\bf r}_a - {\bf r}_\st{cm} \|^2  = {1\over m_\st{tot}}\sum_{a<b}m_a m_b \, r_{ab}^2\,, \label{Icm}
\end{equation}
where $\mathbf r_\st{cm}=  {1\over m_\st{tot}} \sum_a m_a \mathbf r_a$ are the centre-of-mass coordinates and $m_\st{tot} = \sum_a m_a$. We can express $I_\st{cm}$ in terms
of all the remaining variables in the system by solving the constraint $E_\st{tot} = 0$ for it \cite{LongPaper}. This can be done uniquely as we now prove. A well-known analytic result in $N$-body theory is the Lagrange--Jacobi relation \cite{LongPaper}
\begin{equation}
\ddot I_\st{cm}=4E_\st{cm}-2(2+k)V_k\,,\label{ljr}
\end{equation}
which holds for any potential homogeneous of degree $k$. Then for the Newton potential $V_\st{New}$ with $k=-1$ and $E_\st{cm}\ge 0$ it follows that $\ddot I_\st{cm}>0$. This, in turn, means that $I_\st{cm}$ as a function of $t$ is concave upwards, and (half) its time derivative
$D$, the \emph{dilatational momentum}:
\begin{equation}
 D:= \sum_{a=1}^N {\bf r}_a^\st{cm} \cdot {\bf p}^a_\st{cm} \,, ~~~ {\bf p}^a_\st{cm} = {\bf p}^a  - {1\over N} \sum_{b=1}^N {\bf p}^b, \label{dilmtm} 
\end{equation}
is monotonic. The monotonicity of $D$ implies that $I_\st{cm}$ is U-shaped, with a unique minimum,
corresponding to the `turning point' of Fig.~\ref{TriplePicture}. At that point, $D=0$.\footnote{There exist solutions that reach a total collision at which both $I_\st{cm}=0$ and $D=0$, and solutions that lead to parabolic escape at which $I_\st{cm} \to \infty$ and $D \to \infty$~\cite{chenciner1998collisions}. The range of $D$ in these solutions is $[0,\infty)$. These solutions form a measure-zero set with respect to the canonical Louville measure on shape phase space, which is obtained as the quotient of $\mathbb R^{6N}$ (the phase space of $N$ particles in three dimensions) wrt global translations, rotations and dilatations.\label{zm}}

Since $D$ is monotonic, it can be used as a physical time variable $\tau$. Evolution in $\tau$ is generated by a $\tau$-dependent Hamiltonian $\mathcal H$~\cite{LongPaper},
\begin{equation} \label{TimeDepHamiltonian}
 \mathcal H(\tau)=\ln\left( \sum_{a=1}^N {\bm \pi}^a  \cdot {\bm \pi}^a + \tau^2 \right)
 - \ln \left( I_\st{cm}^{\frac 1 2} \, |V_\st{New}| \right) .
\end{equation}
Here ${\bm \pi}^a$ are `shape momenta', defined as
\begin{equation}
{\bm \pi}^a  = \sqrt{\textstyle \frac{I_\st{cm}}{m_a}} \, {\bf p}^a - D \, {\bm \sigma}_a \,,
~~
{\bm \sigma}_a  = \sqrt{\textstyle \frac{m_a}{I_\st{cm}}} \, {\bf r}_a^\st{cm}
\end{equation}
where ${\bm \sigma}_a$ are `shape coordinates', which coordinatize `pre-shape space' $\pshs$, the quotient of configuration space by global translations and dilatations.
One could also quotient by rotations, to obtain `shape space' $\shs$,
the true relational configuration space \cite{LongPaper,FlaviosSDtutorial}. However,
this further quotient is technically impractical and does not affect our
argument (as rotations commute with both the Hamiltonian and the dilatations).
The shape momenta ${\bm \pi}^a $ are the variables canonically conjugate 
to ${\bm \sigma}_a $ \cite{LongPaper}.

Each point of $\shs$ (a shape) is an objective state of the system, freed of unphysical
properties like the overall orientation in absolute space, the position of the center
of mass of the Universe, and the total scale.

The Hamiltonian $\mathcal H(\tau)$ is a function of time $\tau$, the shape degrees of freedom and their conjugate momenta. This allows us to eliminate the evolution of scale from the problem and express the dynamics purely on shape space. The usual description
with scale can be reconstructed from the solution on shape space, for which it is necessary
to specify a nominal initial value of $I_\st{cm}$ (due to dynamical similarity,
this value is completely conventional and unmeasurable).

\section{Complexity}

\begin{figure}[t!]\begin{center}
\includegraphics[width=0.5\textwidth]{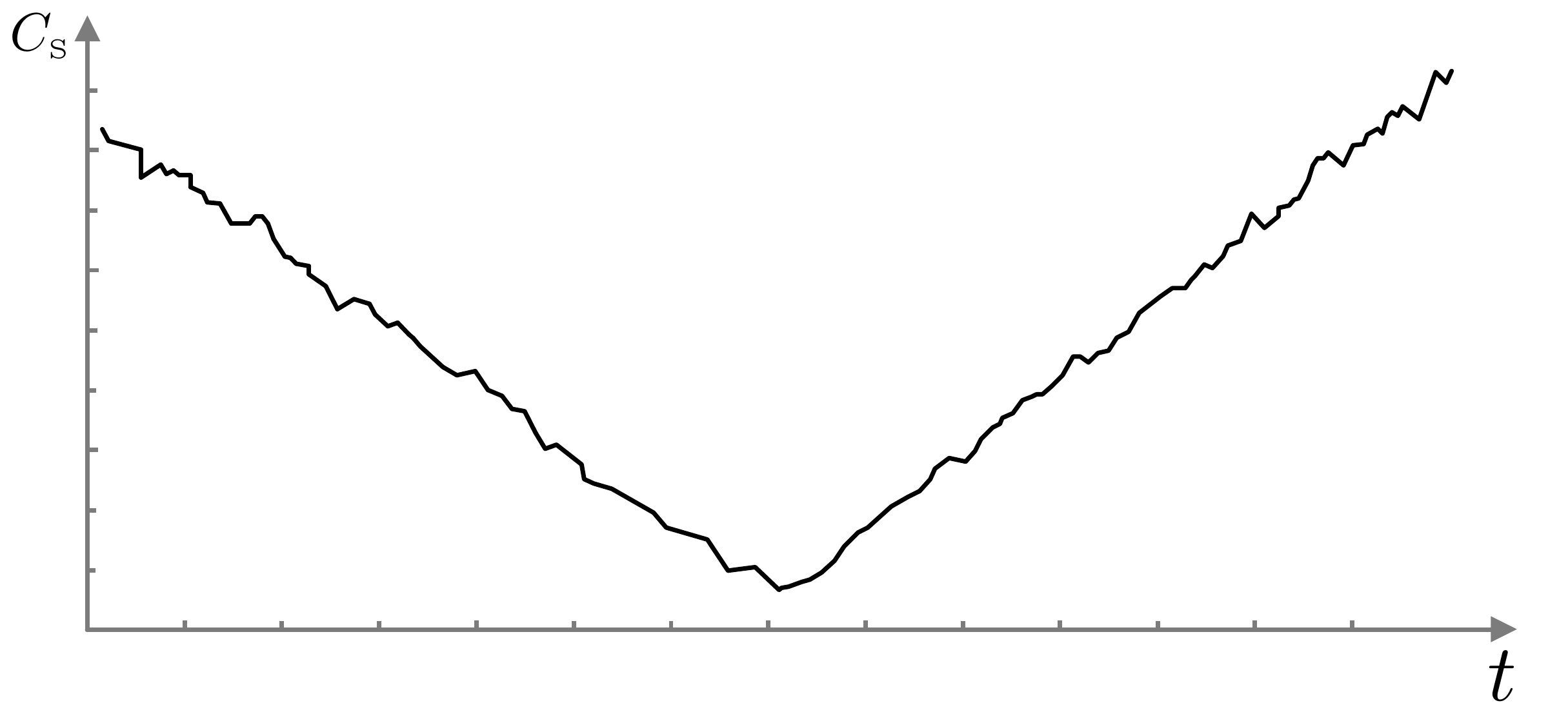}
\caption{ \small Numerical computation of the complexity $C_\st{S}$ vs.~Newtonian time.
This is the typical graph one obtains from a simulation with $N=1000$ particles. There is clear linear growth of $C_\st{S}$ on either side of the central `turning point', where the moment of inertia $I_\st{cm}$ is minimal and the dilatational momentum $D$ vanishes.
\label{1000particles}}
\end{center}
\end{figure}

If we ignore the  nominal scale of the system, what characterizes best its overall state?
It must be some dimensionless measure of inhomogeneity that distinguishes
the central `turning point' shown in Fig.~\ref{TriplePicture} from the states either side. To our knowledge, the objective criterion we now propose is new.
There are two simple lengths that characterize the system. 
One is the \e{root-mean-square length} $\ell_\st{rms}$:
\bq
\ell_\st{rms}:={1\over m_\st{tot}}\sqrt{\sum_{a<b}m_am_b\,r_{ab}^2} = {1\over m_\st{tot}} I_\st{cm}^{\frac 1 2} \,, \label{rms}
\ee
which is dominated by the largest 
$r_{ab}$ and measures the overall size of the system.  The other is the \e{mean harmonic length} $\ell_\st{mhl}$:
\bq
{1\over\ell_\st{mhl}}:={1\over m_\st{tot}^2}{\sum_{r<a}{m_am_b\over r_{ab}}}\label{mhl} =  {1\over m_\st{tot}^2} \left| V_\st{New} \right| \,.
\ee
This length is dominated by the smallest  $r_{ab}$ and measures how close
to each other are the tightest particle pairs. We obtain an observable on shape space
by taking the ratio of these two lengths, $C\st{S}:= \ell_\st{rms}/\ell_\st{mhl}$ and call it the \e{complexity} of the system. It is a good measure of non-uniformity or clustering. Even for relatively small $N$, $\ell_\st{rms}$ (\ref{rms}) changes little if two particles approach each other or even coincide. In contrast, $\ell_\st{mhl}$ (\ref{mhl})   is sensitive to any clustering and tends to zero if that happens. Moreover, while $C_\st{S}$ grows with clustering, Battye et al's  \cite{BattyeGibbons} numerical calculations for the equal-mass case indicate that the minima of $C_\st{S}$ for $N\approx 10^3-10^4$ correspond to extraordinarily uniform (super-Poissonian) shapes. 
Figure~\ref{1000particles} is a plot of $C_\st{S}$ for a typical solution
with $N=1000$: it shows an obvious secular growth away from the point of minimal
expansion.\footnote{The graph showing conjectubehavior of the Universe's entropy in the 
bottom right corner of Fig.~\ref{1000particles} in \cite{CarrollChen} is very like our 
Fig.~\ref{1000particles}. In both cases the most uniform state of the Universe occurs 
symmetrically in the middle.}
We will show below how this can be understood as an intrinsic property of the dynamics on shape space.

Notice that
$ - \ln \left( I_\st{cm}^{\frac 1 2} \,| V_\st{New} | \right) = - \ln C_\st{S} $ in Eq.~(\ref{TimeDepHamiltonian}) plays the role of the potential that attracts the system towards
more inhomogeneous shapes. Also, $C_\st{S}$ has minima and saddle points, but no maxima: it is unbounded
above. This will prove crucial for understanding of the system's behavior.

\section{Scale acts as Friction on Shapes}

The dynamics on shape space is described by a $\tau$-dependent Hamiltonian~(\ref{TimeDepHamiltonian}). Dimensional analysis allows us to eliminate this $\tau$ dependence.
Introduce $\lambda = \log \tau$, and divide the shape momenta
by $D$: ${\bm \omega}^a = {\bm \pi}^a/ D$. These new momenta
${\bm \omega}^a$ are dimensionless. In these new variables
the evolution is generated by a time-independent (autonomous) Hamiltonian~\cite{LongPaper}:
\begin{equation} \label{TimeIndepHam}
H_0 = \log \left( \sum_{a=1}^N {\bm \omega}^a\cdot {\bm \omega}^a + 1 \right) - \log C_\st{S} \,.
\end{equation}
Now, however, the equations of motion,
\begin{equation}
\frac{d {\bm \sigma}_a}{d \lambda} = \frac{\partial H_0}{\partial {\bm \omega}^a},
~~~
\frac{d {\bm \omega}^a}{d \lambda} = - \frac{\partial H_0}{\partial {\bm \sigma}_a} - {\bm \omega}^a ,
\end{equation}
have a non-canonical friction term $- {\bm \omega}^a$ which spontaneously dissipates
the dimensionless momenta ${\bm \omega}^a$.\footnote{Notice that the change of variables
that we performed allows us to describe only {\it half} of each solution: the half before or after $D=0$ \cite{LongPaper}. Each typical solution of the $N$-body problem
maps into two solutions of the equations with friction.}
This and the fact that the potential $-\ln C_\st{S}$ has
no local minima (only infinitely deep potential wells) explain why $C_\st{S}$ grows secularly either side of a unique minimum.

\section{The Arrow of Time}

In the above, we have reformulated the dynamics of the $N$-body problem as
a dynamical system on shape space whose motion is controlled by a potential 
$- \ln C_\st{S}$ and is subject to linear friction. This explains intuitively why 
$C_\st{S}$ and therefore $\ln C_\st{S}$ grows secularly: it is essentially minus the potential energy of a system with friction. Moreover, one can use the results of~\cite{Marchial:1976fi}
to show that typically there is a lower bound on $C_\st{S}$ that grows without bound for large $|\tau|$.\footnote{This result holds also for the atypical solutions that reach $I_\st{cm}=0$. However, we have to assume that at least one bound system forms, and no particle escapes in finite times like in Xia's solution \cite{Xia1992}.}

In the light of this circumstance, it is very natural {\it to identify an arrow of time with the direction in which structure, measured in our case by $C_\st{S}$, grows.} We then have a dynamically-enforced scenario with one past (the minimum of $C_\st{S}$, which occurs near $\tau=0$) and two futures. The growth-of-complexity arrow always points away from the unique past. In the atypical solutions that terminate with $I_\st{cm}=0$ (footnote \ref{zm}) there is one past and only one future \cite{LongPaper}.

\section{Growth of Information}

Our evidence for the passage of time  is in locally stored records (including memory), which, by agreeing with each other, lead us to believe in a dynamical law that has generated them
from a unique past \cite{Barbour1999}.

Complexity is a prerequisite for storage of information
in local subsystems and therefore the formation of records. We now note that a notion of local records can be found in a model as simple as our $E_\st{tot} = {\bf J}_\st{tot} = 0$ $N$-body problem.

Recall that the system, for large $|t|$, breaks up typically into disjoint subsystems  drifting apart linearly in Newtonian time $t$. These subsystems get more and more
isolated, and one can associate dynamically generated local information with them.
For this, we use the result of Marchal and Saari \cite{Marchial:1976fi} that as $t\rightarrow \infty$ each subsystem $\mathcal J$ develops asymptotically
conserved quantities:
\begin{equation}\label{MarchialSaariBounds}
\begin{aligned}
E_{\mathcal J} (t) = E_{\mathcal J}(\infty) + \mathcal O(t^{- 5/3}) ,\\
{\bf J}_{\mathcal J} (t) = {\bf J}_{\mathcal J}(\infty) + \mathcal O(t^{- 2/3}) , \\
{\bf X}_{\mathcal J} (t) / t = {\bf V}_{\mathcal J}(\infty) + \mathcal O(t^{-1/3}).
\end{aligned}
\end{equation}
Here, $E_{\mathcal J}(t), {\bf J}_{\mathcal J} (t) ,{\bf X}_{\mathcal J} (t)$ are, respectively,
the energy, angular momentum, and the distance of the subsystem from the
centre of mass of the total system. The quantities $E_{\mathcal J}(\infty), {\bf J}_{\mathcal J} (\infty) ,{\bf V}_{\mathcal J} (\infty)$ are constants to which $E_{\mathcal J}(t), {\bf J}_{\mathcal J} (t) ,{\bf X}_{\mathcal J} (t)/t$ asymptote.  Let $Y(t)$ be any one of these. Then at any finite time $t$,  the quantity $\left|Y(s)/Y(\infty) -1\right|$ will be smaller than $10^{-\mathcal N (Y,t)}$ for all $s >t$. Here, the integer $\mathcal N (Y,t)$ is a measure of 
how much we know about $Y(\infty)$ from observation of $Y$ up to time $t$ and can be defined as
\begin{equation}
\mathcal N (Y,t) = \left\lfloor \log_{10} \left( \frac {Y(\infty)}{\Delta Y(t)} \right) \right\rfloor \,,
\end{equation}
where $\Delta Y(t) = \max_{s>t}|Y(s)-Y(\infty)|$ means the maximum oscillation (bounded by Eq.~\ref{MarchialSaariBounds}) that the quantity $Y$ can attain after time $t$. Essentially $\mathcal N (Y,t)$ is the number of decimal digits of $Y(\infty)$ that 
we know after time $t$ (with the caveat that in a decimal representation $1 = 0.999999\dots$). It's easy to see how the bounds~(\ref{MarchialSaariBounds}) imply a monotonic
growth of $\mathcal N (Y,t) $. This growth goes along with the growth of $C_\st{S}$, and for the same reason: namely, the subsystems get more and more isolated from each other.

\section{Spontaneous Geometrogenesis}

In the foregoing, we eliminated from the dynamics both Newtonian time $t$ and
scale (represented by $I_\st{cm}$). Although extraneous and superfluous for the
shape dynamics, it is interesting to see how they emerge as {\it effective} concepts.

At late $\tau$'s typical solutions will contain bound systems, like Kepler pairs.
These will be stably bound for a long interval of $\tau$ (many of them forever).
A Kepler pair is characterized by an orbital period, a major axis, and a direction of the  angular momentum. As we just noted, these quantities are almost constant, in the sense that they 
fluctuate around constant values. So each well isolated Kepler pair represents
a physical rod and clock, and a frame of reference (because it provides almost-inertial
axes). The motions of particles near a Kepler pair, measured by the scale provided
by the semimajor axis and the clock provided by the orbital period, are therefore very well approximated by Newton's equations in the frame defined
by the direction of angular momentum and orbital axes.

Thus, in the asymptotic regime, Newtonian physics and a Newtonian framework of space and time in which it holds both emerge spontaneously from a dynamics of shapes with friction.

\section{Extension to General Relativity}

We do not propose the $N$-body model as a phenomenological
model for cosmology; rather it serves as a  toy model to explain a new mechanism. A good cosmological model requires
General Relativity (GR), which, as we will argue in this section, possesses the right properties to give rise to the same mechanism.
However, if the $N$-body problem was interpreted literally as a (coarse-grained) cosmological model (e.g., as a model for the evolution of the distribution of galaxies), one would need to include a positive
cosmological constant. This does not, for the observed value, disrupt bound systems and actually enhances the growth of complexity as we defined it (it increases the $\ell_\st{rms}$ factor without affecting the $\ell_\st{mhl}$ factor significantly).

As we said, the architectonic features of Newtonian gravity that, as we have shown, generate arrows of time, are also present in GR.  GR can be described in a conformally-invariant way through the theory called Shape Dynamics (SD) \cite{gryb:shape_dyn,FlaviosSDtutorial}. This theory involves only conformally-invariant degrees of freedom, plus a single and global scale degree of freedom, which can be identified with the total volume $V$ of a compact spatial slice of constant mean extrinsic curvature (CMC) in spacetime.

In SD, $V$ plays the same role as $I_\st{cm}$ in the $N$-body problem. Its conjugate momentum, known as York time \cite{LongPaper}, is monotonic like to $D$, and can be used as a time variable. In this case too the dynamics is generated by a time-dependent Hamiltonian that can be made time-independent by switching to dimensionless variables and a logarithmic time.

The usual GR description in terms of a spacetime metric can be locally constructed from the conformally-invariant one by solving two elliptic equations, which locally admit unique solutions. Those are the `Lichnerowicz--York equation', giving a local spatial scale (a local version of $I_\st{cm}$) and an equation giving proper time (the lapse of Arnowitt--Deser--Misner gravity)~\cite{LongPaper}.

\section{Conclusions}

Our results are a `proof of principle': {\it all} the solutions of a time-symmetric dynamical law suited to approximate our Universe have a strongly time-asymmetric behavior for internal observers. So far as we know, this conclusion is new. It follows from the exact Lagrange--Jacobi relation (\ref{ljr}) and special properties of the model's potential. Of course, it has long been known that gravity causes clustering of an initially uniform matter distribution. In our Universe, this is reflected above all in the formation of galaxies and is the most striking macroscopic arrow. Our novelty is that in all solutions of the $N$-body problem the dynamical law guarantees, without any past hypothesis, an epoch of relative uniformity out of which growth of structure must be observed. We conclude that the origin of time's arrow is not necessarily to be sought in initial conditions but rather in the structure of the law which governs the Universe.

\section{Acknowledgements}

We thank our referees for many insightful comments that helped to improve substantially this letter.
We also thank Alain Albouy, Niayesh Ashfordi, Alain Chenciner, George Ellis, Brendan Foster, Gary Gibbons, Phillipp H{\"o}hn, Viqar Husain, Christian Marchal, Richard Montgomery and Lee Smolin for discussions. Special thanks to Jerome Barkley for his numerical calculations and permission to show some of his results.
TK was supported in part through NSERC.
Perimeter Institute is supported by the Government of Canada through Industry Canada
and by the Province of Ontario through the Ministry of Economic Development and Innovation. 
This research was also partly supported by grants from FQXi and the John Templeton Foundation.

%

\begin{thebibliography}{100}

\bibitem{zeh2007physical}
H.~D. Zeh, {\em The Physical Basis of the Direction of Time}.
\newblock Springer, 2007.

\bibitem{GibbonsEllis}
G.~W. Gibbons and G.~F.~R. Ellis, ``Discrete Newtonian Cosmology,''
  \href{http://arxiv.org/abs/1308.1852}{{\ttfamily arXiv:1308.1852
  [astro-ph.CO]}}.

\bibitem{barbourbertotti:mach}
J.~Barbour and B.~Bertotti, ``{Mach's principle and the structure of dynamical
  theories},'' {\em Proc. R. Soc. A} {\bfseries 382} no.~1783, (1982) 295--306.

\bibitem{JuliansReview}
J.~Barbour, ``{Shape dynamics. An introduction},'' in {\em Quantum Field Theory
  and Gravity. Proc. Conference at Regensburg 2010}, F.~Finster, ed.
\newblock Birkh\"auser, 2012.
\newblock \href{http://arxiv.org/abs/1105.0183}{{\ttfamily arXiv:1105.0183
  [gr-qc]}}.

\bibitem{FlaviosSDtutorial}
F.~Mercati, ``A Shape Dynamics Tutorial,'' to appear soon on the arXiv.

\bibitem{LongPaper}
J.~Barbour, T.~Koslowski, and F.~Mercati, ``A Gravitational Origin of the
  Arrows of Time,'' \href{http://arxiv.org/abs/1310.5167}{{\ttfamily
  arXiv:1310.5167 [gr-qc]}}.

\bibitem{Marchial:1976fi}
C.~Marchal and D.~Saari, ``On the final evolution of the n-body problem,'' {\em
  J. Diff. Equ.} {\bfseries 20} no.~1, (1976) 150--186.

\bibitem{Landau-Lifshitz}
L.~D. Landau and E.~M. Lifshitz, {\em Course of theoretical physics}, vol.~1:
  Mechanics, ch.~II, sec. 10.
\newblock Pergamon Press, Oxford, 1976.

\bibitem{chenciner1998collisions}
A.~Chenciner, ``Collisions totales, mouvements completement paraboliques et
  reduction des homotheties dans le probleme des n corps.,'' {\em Regular and
  chaotic dynamics} {\bfseries 3} no.~3, (1998) 93--106.

\bibitem{BattyeGibbons}
R.~Battye, G.~Gibbons, and P.~Sutcliffe, ``Central Configurations in Three
  Dimensions,'' \href{http://dx.doi.org/10.1098/rspa.2002.1061}{{\em Proc. R.
  Soc.} {\bfseries A459} (2003) 911--943},
  \href{http://arxiv.org/abs/hep-th/0201101}{{\ttfamily arXiv:hep-th/0201101}}.



\bibitem{CarrollChen}
S.~M. Carroll and J.~Chen, ``Spontaneous inflation and the origin of the arrow
  of time,'' \href{http://arxiv.org/abs/hep-th/0410270}{{\ttfamily
  arXiv:hep-th/0410270 [hep-th]}}.


\bibitem{Xia1992}
Z.~Xia, ``The Existence of Noncollision Singularities in Newtonian Systems,''
  {\em Ann. Math.} {\bfseries 135} (1992) 411--468.

\bibitem{Barbour1999}
J.~Barbour, ``The end of time: The next revolution in physics,''. Oxford
  University Press, UK.

\bibitem{gryb:shape_dyn}
H.~Gomes, S.~Gryb, and T.~Koslowski, ``{Einstein gravity as a 3D conformally
  invariant theory},''
  \href{http://dx.doi.org/10.1088/0264-9381/28/4/045005}{{\em Class. Quant.
  Grav.} {\bfseries 28} (2011) 045005},
  \href{http://arxiv.org/abs/1010.2481}{{\ttfamily arXiv:1010.2481 [gr-qc]}}.

\end{thebibliography}

\end{document}